\documentstyle[times,pramana,epsf,floats]{ias}

\newcommand{\n}{\noindent}
\newcommand{\be}{\begin{equation}}
\newcommand{\ee}{\end{equation}}
\newcommand{\bearr}{\begin{eqnarray}}
\newcommand{\eearr}{\end{eqnarray}}
\newcommand{\bmath}{\begin{mathletters}}
\newcommand{\emath}{\end{mathletters}}

\begin{document}
\title{Entanglement production in Quantized Chaotic Systems}
\author{Jayendra N. Bandyopadhyay$^a$ and Arul Lakshminarayan$^b$}
\address{$^a$Physical Research Laboratory, Navrangpura,
Ahmedabad 380009, India.\\
$^b$Department of Physics, Indian Institute of Technology, Madras, 
Chennai 600036, India}
\abstract{Quantum chaos is a subject whose major goal is to identify and to
investigate different quantum signatures of classical chaos. Here we study 
entanglement production in coupled chaotic systems as a possible quantum 
indicator of classical chaos. 
We use coupled kicked tops as a model for our extensive numerical studies. We 
find that, in general, presence of chaos in the system produces more 
entanglement. However, coupling strength between two subsystems is also very 
important parameter for the entanglement production. Here we show how chaos can
lead to large entanglement which is universal and describable by random matrix 
theory (RMT). We also explain entanglement production in coupled strongly 
chaotic systems by deriving a formula based on RMT. This formula is valid for 
arbitrary coupling strengths, as well as for sufficiently long time. Here
we investigate also the effect of chaos on the entanglement 
production for the mixed initial state. We find that many properties of the 
mixed state entanglement production are qualitatively similar to the pure 
state entanglement production. We however still lack an analytical 
understanding of the mixed state entanglement production in chaotic systems.}

\maketitle
\section{Introduction}

Entanglement is a unique quantum phenomenon which can be observed in a system 
consists of at least two subsystems. In case of an entangled system even if we 
know the exact state of the system, it is not possible to assign any pure state
to the subsystems and that leads to the well-known unique quantum correlations
which exists even in spatially well separated pairs of subsystems. This 
phenomenon was first discussed by Schr\"{o}dinger to point out the 
nonclassicality implied by the quantum mechanical laws \cite{schro}. This
remarkable feature of quantum mechanics has recently been identified as a 
resource in many areas of quantum information theory including quantum 
teleportation \cite{bennett1}, superdense coding \cite{bennett2} and quantum
key distribution \cite{ekert}. Moreover, entanglement is also a key ingredient
of all the proposed quantum algorithms which outperform their classical
counterparts \cite{shor,grover}.   

Quantum mechanical study of classically chaotic systems is the subject matter
of `quantum chaos' \cite{haake,stockman}. A major challenge of quantum chaos
is to identify quantum signatures of classical chaos. Various signatures have
been identified, such as the spectral properties of the generating Hamiltonian
\cite{bohigas}, phase space scarring \cite{heller}, hypersensitivity to
perturbation \cite{caves}, and fidelity decay \cite{peres}, which indicate
presence of chaos in underlying classical system. Recent studies have shown
that entanglement in chaotic systems can also be a good indicator of the
regular to chaotic transition in its classical counterpart
\cite{furuya,sarkar,arul,our1,our2,tanaka,arul_subra,lahiri,scott,santos,wang,kus}. 
A study of the connections between chaos and entanglement is interesting 
because the two phenomena are {\it prima facie} uniquely classical and quantum,
respectively. This is definitely an important reason to study entanglement in 
chaotic systems. Moreover, presence of chaos has also been identified in some 
realistic model of quantum computers \cite{shepel}. 

In this paper, we have investigated entanglement production in coupled chaotic
systems. We have used coupled kicked tops as a model for our whole study. We 
have considered the entanglement production for both chaotic and regular cases.
Moreover, we have also considered the effect of different coupling strengths
on entanglement production. Most of the earlier studies have considered the 
effect of chaos on entanglement production for the case of initially pure state
of the overall system. A basic assumption of these studies is that the initial 
state of the overall system is completely known. However, in many of the 
realistic scenarios, we do not have a complete knowledge of the state of a 
quantum system. For instance, when a quantum system interacts with its 
surroundings, it is not possible to know the exact state of the system. We may 
only express the state of the system as {\it statistical} mixture of different 
pure states, and that is a {\it mixed} state. In this paper we have also 
studied mixed state entanglement production in chaotic systems. 

This paper is organized as follows. In the next section we discuss about 
classical and quantum properties of two coupled kicked tops, our primary 
model. Then we have
defined the measures of both pure and mixed state entanglement. Finally, we 
have concluded this section with a discussion on the initial states (both
pure and mixed) used here. In Sec.\ref{sec3}, we present the numerical 
results on the entanglement production in coupled kicked tops for different
single top dynamics and also for different coupling strengths. Here we have 
studied entanglement production for both pure and mixed initial state. In
Sec.\ref{sec4}, we derive the statistical universal bound on 
entanglement using random matrix theory (RMT). We also derive an 
approximate formula, based on RMT, to explain the entanglement production in
coupled strongly chaotic systems. Finally, we summarize in Sec.\ref{sec5}  

\section{{\label{sec2}}Preliminaries}

\subsection{Coupled kicked tops}
\subsubsection{Quantum top}

The single kicked top is characterized by an angular momentum vector ${\bf J}
= (J_x, J_y, J_z)$, where these components obey the usual commutation
rules. The Hamiltonian of the single top is given by \cite{haake_top}
\be
H(t) = \frac{\pi}{2} J_y + \frac{k}{2j} J_{z}^{2} \sum_{n = - \infty}
^{n = + \infty}\, \delta(t - n).
\ee
\n The first term describes free precession of the top around $y$ axis with
angular frequency $\pi/2$, and the second term is due to periodic 
$\delta$-function kicks. The second term is torsion about $z$ axis by an angle
proportional to $J_z$, and the proportionality factor is a dimensionless 
constant $k/2j$. Now the Hamiltonian of the coupled kicked tops can be written,
following Ref. \cite{sarkar}, as 
\be
{\cal H}(t) = H_1(t) + H_2(t) + H_{12}(t),
\ee
\n where  
\bmath
\bearr
H_i(t) &\equiv & \frac{\pi}{2} J_{y_i} + \frac{k_1}{2j} J_{z_i}^{2} \sum_n
\,\delta(t - n)\\
H_{12}(t) &\equiv & \frac{\epsilon}{j} J_{z_1} J_{z_2} \sum_n\,\delta(t - n),
\eearr
\emath
\n where $i = 1,2$. Here $H_i(t)$'s represent the Hamiltonians of the 
individual tops, and $H_{12}(t)$ is the coupling between the tops via 
spin-spin interaction with a coupling strength $\epsilon/j$. Corresponding
time evolution operator, defined in between two consecutive kicks, is given
by
\be
U_T = U_{12}^{\epsilon}(U_1 \otimes U_2) = U_{12}^{\epsilon}
[(U_{1}^{k} U_{1}^{f}) \otimes (U_{2}^{k} U_{2}^{f})],
\ee
\n where the different terms are given by,
\be
\hspace{-1cm}
U_{i}^{f}\equiv\exp\left(-i\frac{\pi}{2}J_{y_i}\right),
\,U_{i}^{k}\equiv\exp\left(-i\frac{k}{2j}J_{z_i}^{2}\right),\,
U_{12}^{\epsilon}\equiv\exp\left(-i\frac{\epsilon}{j}J_{z_1} J_{z_2}
\right)
\ee
\n and as usual $i=1,2$.

\subsubsection{Classical top}

The classical map corresponding to the coupled kicked tops can be obtained 
from the quantum description with the Heisenberg picture in which the
angular momentum operators evolve as 
\be
{\bf J}_{n+1} = U_{T}^{\dagger}{\bf J}_n U_T.
\ee
\n Explicit form of this angular momentum evolution equation for each component
of the angular momentum is presented in Ref.\cite{our2}. We now proceed 
by rescaling the angular momentum operator as $(X_i, Y_i, Z_i) \equiv (J_{x_i},
J_{y_i}, J_{z_i})/j$, for $i=1,2$. The commutation relations satisfied by the
components of this rescaled angular momentum vector as follow : $[X_i, Y_i]
= i Z_i/j, [Y_i, Z_i] = i X_i/j$ and $[Z_i, X_i] = i Y_i/j$. Therefore, in
$j \rightarrow \infty$ limit, components of this rescaled angular momentum
vector will commute and become classical $c$-number variables. In this 
large-$j$ limit, we obtain the classical map corresponding to coupled kicked 
top as \cite{our2}:
\begin{figure}[t]
\epsfxsize=8cm
\centerline{\epsfbox{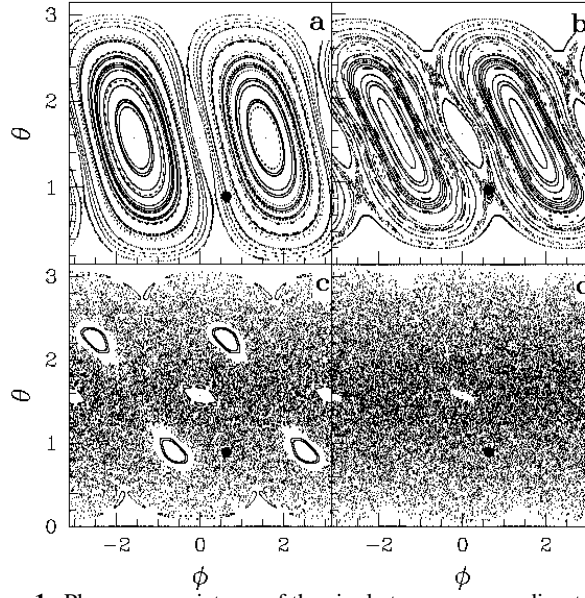}}
\caption{Phase space pictures of the single top, corresponding to different
parameter values, are presented. (a) $k = 1.0$. Phase space is mostly covered
by the regular region. (b) $k = 2.0$. The phase space is still very much
regular, but now a thin stochastic layer can be observed at the separatrix.
(c) $k = 3.0$. The phase space is truly mixed type. Few regular elliptic
islands are visible inside the chaotic region. (d) $k = 6.0$. The phase space
is almost covered by the chaotic region with few tiny elliptic islands. The
solid circle $(\bullet)$ is the point at which we will construct the initial
wave packet during our study of the pure state entanglement production }
\label{fig1}
\end{figure}
\bmath
\bearr
X_{1}^{\prime} &=& Z_1 \cos \Delta_{12} + Y_1 \sin \Delta_{12},\\
Y_{1}^{\prime} &=& - Z_1 \sin \Delta_{12} + Y_1 \cos \Delta_{12},\\
Z_{1}^{\prime} &=& - X_1,\\
X_{2}^{\prime} &=& Z_2 \cos \Delta_{21} + Y_2 \sin \Delta_{21},\\
Y_{2}^{\prime} &=& - Z_2 \sin \Delta_{21} + Y_2 \cos \Delta_{21},\\
Z_{2}^{\prime} &=& - X_2,
\eearr
\emath
\n where
\be
\Delta_{12} \equiv k X_1 + \epsilon X_2 ~~\mbox{and}~~ \Delta_{21}
\equiv k X_2 + \epsilon X_1.
\ee
In the limit $\epsilon \rightarrow 0$, the classical map for the coupled kicked 
tops decouple into the classical map for two single tops. The classical map
for one such uncoupled top can be written as
\bmath
\bearr
X^{\prime} &=& Z \cos kX + Y \sin kX \\
Y^{\prime} &=& -Z \sin kX + Y \cos kX \\
Z^{\prime} &=& -X.
\eearr
\emath
\n It is clear from the above expression that the variables $(X, Y, Z)$ lie
on the sphere of radius unity, i.e. $X^2 + Y^2 + Z^2 = 1$. Consequently, we can
parameterize the dynamical variables in terms of the polar angle $\theta$ and
the azimuthal angle $\phi$ as $X = \sin \theta\,\cos \phi, Y = \sin \theta\,
\sin \phi,$ and $Z = \cos \theta$. In Fig. \ref{fig1}, we have presented the
phase space diagrams of the single top for different values of the parameter
$k$. For $k = 1.0$, as shown in Fig.\ref{fig1}(a), the phase space is mostly 
covered by regular orbits, without any visible stochastic region. Our initial
wave packet, marked by a solid circle at the coordinate $(0.89,0.63)$,
is on the regular elliptic orbits. As we further increase the
parameter, regular region becomes smaller.  Fig.\ref{fig1}(b) is
showing the phase space for $k = 2.0$. Still the phase space is mostly
covered by the regular region, but now we can observe a thin
stochastic layer at the separatrix. In this case, the initial
wave packet is on the separatrix. For the change in the parameter value
from $k=2.0$ to $k=3.0$, there is significant change in the phase
space. At $k=3.0$, shown in Fig.\ref{fig1}(c), the phase space is of a
truly mixed type. The size of the chaotic region is now very large
with few regular islands. At this parameter value, the initial
wave packet is inside the chaotic region. Fig.\ref{fig1}(d) is showing
the phase space for $k=6.0$. Now the phase space is mostly covered by
the chaotic region, with very tiny regular islands. Naturally, our
initial wave packet is in the chaotic region. 

\subsection{Measures of entanglement}
\subsubsection{Pure state}

Entanglement measure for a system consisting of two subsystems (bipartite) is 
well defined if overall state of the system is in a pure state. In this case 
subsystem von Neumann entropy, i.e. von Neumann entropy of the reduced density 
matrices (RDMs), is a measure of entanglement. If there is no 
entanglement among the two subsystems, then the RDMs will correspond to 
density matrices of pure states and hence the subsystem von Neumann entropy 
will vanish. Otherwise, in case of entanglement, a non-zero value of the 
subsystem von Neumann entropy will be a measure of entanglement among the two
subsystems.

Let us assume that the state space of a bipartite quantum system is ${\cal H}
= {\cal H}_1 \otimes {\cal H}_2$, where $\mbox{dim}\,{\cal H}_1 = N \leq 
\mbox{dim}\,{\cal H}_2 = M$, and $\mbox{dim}\,{\cal H} = d = N M$. If $\rho =
\sum_i p_i | \phi_i\,\rangle \langle \phi_i |$ is an ensemble representation
of an arbitrary state in ${\cal H}$, the entanglement of formation is found
by minimizing $\sum_i p_i E\left(|\phi_i\,\rangle\right)$ over all possible
ensemble realizations. Here $E$ is the von Neumann entropy of the RDM of the
state $|\phi_i\rangle$ belonging to the ensemble, i.e. its entanglement. For
pure states $|\psi\rangle$ there is only one unique term in the ensemble 
representation and the entanglement of formation is simply the von Neumann 
entropy of the RDM.

The two RDMs of any bipartite pure state $|\psi\rangle$ are $\rho_1 = 
\mbox{Tr}_2 (|\psi\rangle\langle\psi |)$ and $\rho_2 = \mbox{Tr}_1 
(|\psi\rangle\langle\psi |)$. The Schmidt decomposition of $|\psi\rangle$ is 
the optimal representation in terms of a product basis and is given by
\be
|\psi\rangle = \sum_{i=1}^{N}\,\sqrt{\lambda}_i |\phi_{i}^{(1)}\rangle 
|\phi_{i}^{(2)}\rangle,
\ee
\n where $0 < \lambda_i \leq 1$ are the (non-zero) eigenvalues of either RDMs
and the vectors are the corresponding eigenvectors. The von Neumann entropy
$S_V$ is the entanglement $E(|\psi\rangle)$ is given by
\be
S_V = - \mbox{Tr}_l (\rho_l \ln \rho_l) = - \sum_{i=1}^{N}\,\lambda_i 
\ln(\lambda_i)\,\,;\,\,l = 1, 2.
\label{Sv}
\ee

The von Neumann entropy can only be calculated in the eigenbasis of the RDMs
due to the presence of {\it logarithmic} function in its definition.
Therefore it is not easy to calculate this measure unless one has some
information of the eigenvalues of the RDMs. Consequently linearized version of 
the von Neumann entropy, called linear entropy, has also become a popular 
measure of entanglement. This measure of entanglement is defined as
\be
S_R = 1 - \mbox{Tr}_l \rho_{l}^{2} = 1 - \sum_{i=1}^{N}\,\lambda_{i}^{2}\,\,;
~l=1,2.
\ee
\n The linear entropy can be calculated without knowing the eigenvalues of the
RDMs, because $\mbox{Tr}_l \rho_{l}^{2}$ is equal to the summation of
absolute square of all the elements of RDMs. However, strictly speaking, the
linear entropy is not a true measure of entanglement, rather it is a measure
of {\it mixedness} of the subsystems which increases with entanglement among
the two subsystems. Therefore, the linear entropy can be considered as an
approximate measure of entanglement.       

\subsubsection{Mixed state}

A major issue related to the study of the mixed state 
entanglement is lack of unique measure of entanglement. Probably this issue
has discouraged any work related to the mixed state entanglement production
in chaotic systems. Recently {\it Vidal and Werner} \cite{wern_vid} have
proposed a computable measure of entanglement called {\it Log-negativity}
following Peres' criterion of separability \cite{peres1}. We use this measure
to characterize mixed state entanglement production in chaotic systems. Basic
idea of this measure is very simple and straightforward to state.

A most general form of a separable bipartite mixed state is given by
\be
\rho = \sum_i\,p_i\,\rho_{i}^{(1)} \otimes \rho_{i}^{(2)},
\ee
\n where the positive weight factors $p_i$ satisfy $\sum_i\,p_i = 1$, 
$\rho_{i}^{(1)}$ and $\rho_{i}^{(2)}$ are density matrices for the two
subsystems. We can construct a matrix $\rho^{T_2}$ from $\rho$ by taking 
transpose only over second subspace, i.e.
\be
\rho^{T_2} = \sum_i\,p_i\,\rho_{i}^{(1)} \otimes \left(\rho_{i}^{(2)}\right)^T.
\ee  
\n This partial transpose operation is definitely not a unitary operation, but 
$\rho^{T_2}$ is still Hermitian. The transposed matrices $\left(\rho_{i}^{(2)}
\right)^T$ are positive matrices, and hence they are legitimate density 
matrices. Consequently, if $\rho$ is separable, $\rho^{T_2}$ is a positive 
matrix. This is also true for $\rho^{T_1}$. In general, this is a necessary 
condition of separability.  

Log-negativity measures the degree to which $\,\rho^{T_2}\,\left(\mbox{or}\,
\rho^{T_1}\right)$ fails to be positive. If $\rho$ is an entangled state,
then $\rho^{T_2}$ may have some negative eigenvalues. The Log-negativity is 
logarithm of the sum of absolute value of the negative eigenvalues of 
$\rho^{T_2}$ which vanishes for unentangled state. It can be shown by simple 
algebraic manipulation that the sum of absolute value of all the negative
eigenvalues of $\rho^{T_2}$ is linearly related to the sum of absolute value 
of all the eigenvalues of $\rho^{T_2}$. Therefore, the Log-negativity measure
$E_N(\rho)$ can be defined as
\be     
E_N(\rho) = \ln \left(\sum_{i=1}^{d} |\lambda_i| \right)
\ee
\n where $d$ is the dimension of $\rho$.

\subsection{Initial state}

\subsubsection{pure state} 

We use generalized SU$(2)$ coherent state or the directed angular momentum 
state \cite{haake,haake_top} as our initial state for the individual tops
and this state is given in standard angular momentum basis $|j, m\rangle$ as
\be
\langle j, m | \theta_0, \phi_0\rangle = (1 + |\gamma|^2)^{-j} \gamma^{j-m}
\,\sqrt{\left(\begin{array}{c} 2j\\j + m \end{array}\right)},
\label{coh_state}
\ee
\n where $\gamma \equiv \exp(i\phi_0)\tan(\theta_0/2)$. For the coupled kicked
top, we take the initial state as the tensor product of the directed angular
momentum state corresponding to individual top, i.e.,
\be 
|\psi(0)\rangle = |\theta_{0}^{(1)}, \phi_{0}^{(1)}\rangle\, |\theta_{0}^{(2)}, 
\phi_{0}^{(2)}\rangle,
\ee
\n where $(\theta_{0}^{(i)}, \phi_{0}^{(i)}) = (0.89, 0.63)$ for $i=1,2$.
This initial state is evolved under $U_T$ as $|\psi(n)\rangle = U_{T}^{n}
|\psi(0)\rangle$ for different values of the parameter $k$ and for different 
coupling strength $\epsilon$, and the results are displayed in Fig. \ref{fig2}.

\subsubsection{mixed state}

In this case we have considered a very simple unentangled mixed state, where
the initial state corresponding to first top is mixed and the same 
corresponding to the second top is pure. Mathematically we express this state 
as $\rho(0) = \rho_{1}^{}(0)\,\otimes\,| \psi_2 (0)\rangle\,\langle
\psi_2(0) |$, where $\rho_{1}^{}(0)$ is the initial mixed state of the first 
subsystem and $| \psi_2 (0)\rangle$ is the initial pure state of the second 
subsystem. We take $| \psi_2(0)\rangle$ as a generalized $SU(2)$ coherent 
state as presented above in Eq.(\ref{coh_state}). The mixed state $\rho_1(0)$ 
is a combination of two such coherent states placed at two different points 
on the phase space, i.e.,
\be
\rho_1(0) = p\,| \theta_{a0}^{(1)},\phi_{a0}^{(1)} \rangle\,\langle
\theta_{a0}^{(1)},\phi_{a0}^{(1)} | + (1-p)\,| \theta_{b0}^{(1)},
\phi_{b0}^{(1)} \rangle\,\langle \theta_{b0}^{(1)},\phi_{b0}^{(1)} |.
\label{ini_mix}
\ee
\n Here we choose $\bigl(\theta_{a0}^{(1)},\phi_{a0}^{(1)}\bigr)=(0.89, 0.63)$ 
and $\bigl(\theta_{b0}^{(1)}, \phi_{b0}^{(1)} \bigr) = (2.25, - 0.63)$ in such
a way that the dynamical properties of these points are similar for any value
of $k$. For the second top, we choose $| \psi_2(0) \rangle = 
| \theta_{0}^{(2)} = 0.89, \phi_{0}^{(2)} = 0.63 \rangle$. We only consider 
$p = 1/2$ case, this means the contribution of each coherent state is 
same on the formation of $\rho_1(0)$. The initial mixed state $\rho(0)$ is 
evolved under $U_T$ as $\rho(n) = U_{T}^{n}\rho(0) U_{T}^{- n}$. We study the 
time-evolution of the Log-negativity measure for different $k$ and $\epsilon$, 
and the results are displayed in Fig. \ref{fig3}.

\section{\label{sec3}Numerical results}

\subsection{\label{bounds}Pure state entanglement production}

In Fig.\ref{fig2}, we have presented our results for the entanglement 
production in coupled kicked tops for the spin $j=80$. As we go from top to
bottom window, coupling strength is decreasing by a factor of ten. Top window
corresponds to $\epsilon=10^{-2}$, middle window is showing the results for
$\epsilon=10^{-3}$, and the bottom one corresponds to the case $\epsilon
=10^{-4}$. For each coupling strengths, we have studied entanglement production
for four different single top parameter values, whose corresponding classical 
phase space picture has already been shown in Fig.\ref{fig1}.

\subsubsection{Coupling $\epsilon = 10^{-2}$}

\begin{figure}[t]
\epsfxsize=7.5cm
\centerline{\epsfbox{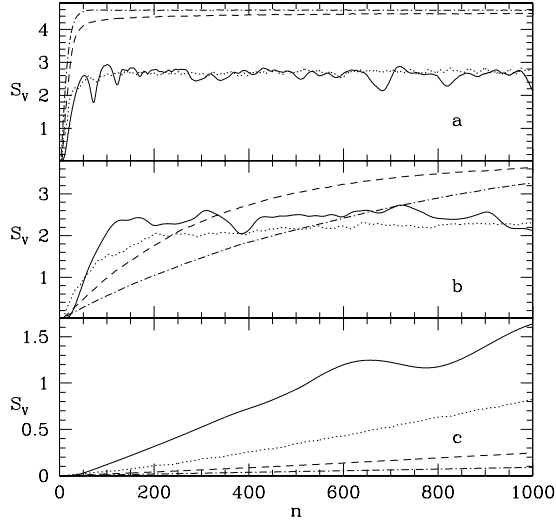}}
\caption{Time evolution of the von Neumann entropy in coupled
kicked tops is presented for different coupling strengths and for different
underlying classical dynamics. (a) $\epsilon = 10^{-2}$. (b) $\epsilon =
10^{-3}$. (c) $\epsilon = 10^{-4}$. Solid line represents $k = 1.0$, dotted line
corresponds to $k = 2.0$, dashed line is for $k = 3.0$ and dash-dot line
represents $k = 6.0$.}
\label{fig2}
\end{figure}

The entanglement production for this strong coupling strength has been 
presented in Fig.\ref{fig2}(a). It shows that there exists a saturation of
$S_V$ for the regular cases ($k=1.0$ and $k=2.0$), which are much less than
the saturation value corresponding to strongly chaotic cases such as when
$k=6.0$. The saturation value of $S_V$ for $k=6.0$ is a statistical bound
$S_V = \ln(N) - \frac{1}{2} \simeq 4.57$ (where $N = 2j + 1 = 161$), which
can be estimated analytically from RMT \cite{our1}, 
and we will discuss about this in the next section. However for $k=3.0$, 
corresponding to a mixed classical phase space, the saturation value of $S_V$
is less than the above mentioned statistical bounds, which indicates the 
influence of the regular regions. These distinct behaviors of the entanglement
saturation can be understood from the underlying classical dynamics. For
$k=1.0$, the initial unentangled state is the product of the coherent 
wave packet placed on some elliptic orbits of each top. Therefore, the 
evolution of this unentangled state under the coupled top unitary operators
is restricted by those elliptic orbits. Finally, the wave packet spreads 
all over those elliptic orbits and the entanglement production reaches its 
saturation value. At $k = 2.0$, the center of the initial coherent state is
inside the separatrix. Therefore, in its time evolution, the spreading of the
wave packet is restricted to be inside the separatrix region. Finally it
spread over the whole separatrix region, and the entanglement production 
arrives at its saturation. At $k = 3.0$ and $k = 6.0$, the initial wave packets
are inside the chaotic region. However, due to the smaller size of the chaotic
region corresponding to the case of $k = 3.0$ than the case corresponding to
$k = 6.0$, the wave packet can spread over less of the phase space for $k = 3.0$
than $k = 6.0$. Consequently, the saturation value of the entanglement 
production is less for $k = 3.0$ than $k = 6.0$.     

\subsubsection{Coupling $\epsilon = 10^{-3}$}

Let us now discuss the case of coupling strength $\epsilon = 10^{-3}$, whose
results are presented in Fig.\ref{fig2}(b). For the non-chaotic cases ($k = 1.0$
and $k = 2.0$), the saturation value of the entanglement production is less
than the entanglement saturation value observed in the stronger coupling case
$(\epsilon = 10^{-2})$. This is because, for weaker coupling case, the 
interaction between two subsystems is less and the individual subsystems 
behave more like isolated quantum systems. Similarly, for the strong chaos 
case $(k = 6.0)$, the entanglement production is well short of the known 
statistical bound $\ln(N) - \frac{1}{2}$.

\subsubsection{Coupling $\epsilon = 10^{-4}$}

The entanglement production for this very weak coupling strength has been 
presented in Fig.\ref{fig2}(c). The entanglement production for the weakly
coupled strongly chaotic system has recently been explained by perturbation
theory \cite{tanaka}. However, the formula presented in that work is only 
valid for short time. In the next section we have presented an approximate
formula for the entanglement production in coupled strongly chaotic systems
which is valid for sufficiently long time and for any arbitrary coupling
strengths. This formula explains the entanglement production for the strongly
chaotic case $(k = 6.0)$. Here we have observed an interesting phenomenon that
the entanglement production is much larger for the non-chaotic cases than the
chaotic cases. Rather, we can say that, for weakly coupled cases, the presence
of chaos in the systems actually suppresses entanglement production.  

\subsection{Mixed state entanglement production}

In Fig.\ref{fig3}, we have presented the Log-negativity measure $E_N(\rho)$ 
of the mixed state entanglement production for different individual top 
dynamics $(k = 1.0, 2.0, 3.0,$ and $6.0)$ and for different coupling strengths. 

\subsubsection{Coupling $\epsilon = 1.0$}

\begin{figure}[t]
\epsfxsize=7.5cm
\centerline{\epsfbox{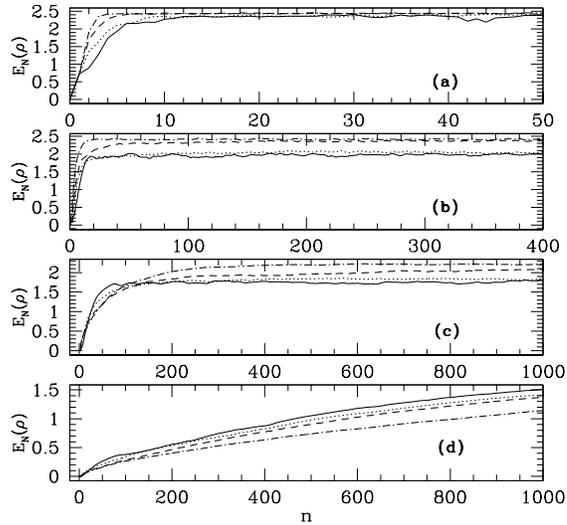}}
\caption{Evolution of the Log-negativity measure (evolving under the
coupled tops time evolution operator $U_T$). Solid lines and dotted lines are
representing the results corresponding to the non-chaotic cases ($k = 1.0$
and $k = 2.0$, respectively). Dashed lines are representing the mixed case
$(k = 3.0 )$ and dash-dot lines are showing the results for the strongly
chaotic case $(k = 6.0 )$. (a) Representing the results for the stronger
coupling strength $(\epsilon = 1.0 )$. (b) This window is showing the
results for $\epsilon = 0.1$. (c) This window is for $\epsilon = 0.01$
case. (d) This window is showing the results for the weak coupling case
$(\epsilon = 0.001$).}
\label{fig3}
\end{figure}

Let us start the discussion with the case of strong coupling $\epsilon = 1.0$, 
whose results are presented in Fig.\ref{fig3}(a). This coupling strength is so 
strong that, irrespective of the individual top dynamics, the overall coupled 
system is chaotic. Therefore, the location of the initial state and the 
dynamics of the individual tops are irrelevant for the saturation of 
$E_N(\rho)$. Consequently, we have observed almost same saturation value 
of $E_N(\rho)$ for all the different individual top dynamics. 
\subsubsection{Coupling $\epsilon = 0.1$}

The time evolution of $E_N(\rho)$ corresponding to $\epsilon = 0.1$ is
presented in Fig.\ref{fig3}(b). For this coupling strength, we have observed 
that the saturation value of $E_N(\rho)$ for the non-chaotic cases 
$(k = 1.0\,\mbox{and}\,k = 2.0)$ are less than the saturation value 
corresponding to other two cases. These lower saturation values of 
$E_N(\rho)$ for the non-chaotic cases indicate the influence of the regular 
orbits on the mixed state entanglement production. We have also noticed for 
the non-chaotic cases that the saturation value of $E_N(\rho)$ is less than 
the saturation value observed in the stronger coupling case $(\epsilon = 1.0)$.
However, the saturation value of $E_N(\rho)$ corresponding to other two cases, 
$k = 3.0$ and $k = 6.0$, are almost equal to the previous case $(\epsilon = 
1.0)$.

\subsubsection{Coupling $\epsilon = 10^{-2}$}

The mixed state entanglement production for the coupling strength 
$\epsilon = 10^{-2}$ has been presented in Fig.\ref{fig3}(c). Here again 
the saturation value of $E_N(\rho)$ corresponding to the non-chaotic cases 
are less than the other two cases. Moreover, due to the weaker coupling,
the saturation value of $E_N(\rho)$ for the non-chaotic cases are less than 
the saturation value observed in the previous two cases of stronger coupling 
strengths $(\epsilon = 1.0$ and $\epsilon = 0.1)$. Here we have first time
observed an interesting phenomenon that the magnitude of $E_N(\rho)$
corresponding to the non-chaotic cases are larger than the other two cases at
least within a time interval $50 \lesssim n \lesssim 100$. However, the 
saturation value corresponding to the case of mixed phase space $(k = 3.0)$ 
is still less than the saturation value corresponding to the chaotic case 
$(k = 6.0)$. 

\subsubsection{Coupling $\epsilon = 10^{-3}$}

Finally, for the weak coupling strength $\epsilon = 10^{-3}$, we have observed
completely different behaviors of the evolution of $E_N(\rho)$ and these
results are presented in Fig.\ref{fig3}(d). For this coupling strength,
$E_N(\rho)$ corresponding to the non-chaotic cases are always higher than the 
chaotic cases within our time of observation. These results imply that, in 
case of weak coupling, the presence of chaos in the system actually suppresses 
the entanglement production. This suppression of the entanglement production by
chaos for the weak coupling case was also observed for the pure state 
entanglement production, which we have already discussed in the previous 
section.

\section{\label{sec4}Some analytical results}

\subsection{Random matrix estimation of the statistical bound on entanglement}

In our numerical study of pure state entanglement production, we have observed
a statistical bound on entanglement for the strongly coupled strongly chaotic
top. An identical property has also been observed for the stationary states of 
coupled standard map \cite{arul} which indicates its universality. The 
parameter values considered in our numerical study, the nearest neighbor 
spacing distribution of the eigenangles of $U_T$ is Wigner distributed, which 
is typical of of any quantized chaotic systems \cite{haake,stockman}. 
Therefore, it is quite reasonable to expect that the statistical bound on 
entanglement can be estimated by random matrix modeling. 
\begin{figure}[t]
\epsfxsize=7.5cm
\centerline{\epsfbox{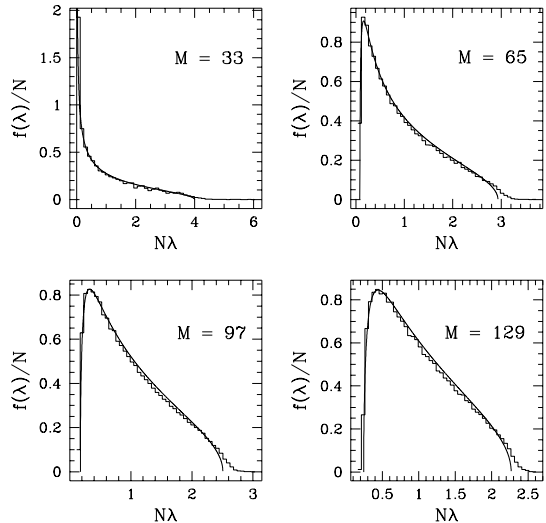}}
\caption{Distribution of the eigenvalues of the RDMs of coupled kicked tops,
averaged over all the eigenstates ($N = 2j_1 + 1 = 33$). Solid curves
correspond to the theoretical distribution function Eq.(\ref{distri}).}
\label{fig4}
\end{figure}
The two RDMs, corresponding to two subsystems, have the structure $A^{\dagger}A$
and $AA^{\dagger}$, where $A$ is a rectangular $N \times M$ matrix containing
the vector components of the bipartite system. We have pointed out the 
distribution of the eigenvalues of these RDMs as \cite{our1}
\bearr   
f(\lambda) &=& \frac{NQ}{2\pi} \frac{\sqrt{( \lambda_{\mbox{\small max}} - 
\lambda )( \lambda - \lambda_{\mbox{\small min}} )}}{\lambda} \nonumber\\
\lambda_{\mbox{\small min}}^{\mbox{\small max}} &=& \frac{1}{N} 
\left( 1 + \frac{1}{Q} \pm \frac{2}{\sqrt{Q}} \right),
\label{distri}
\eearr
\n where $\lambda \in [\lambda_{\mbox{\small min}}, \lambda_{\mbox{\small max}}
], \,Q=M/N,$ and $N f(\lambda)\,d\lambda$ is the number of eigenvalues within
$\lambda$ to $\lambda + d\lambda$. This has been derived under the assumption
that both $M$ and $N$ are large. Note that this predicts a range of eigenvalues
for the RDMs that are of the order of $1/N$. For $Q \ne 1$, the eigenvalues 
of the RDMs are bounded away from the origin, while for $Q=1$ there is a
divergence at the origin. All of these predictions are seen to be borne
out in numerical work with coupled tops.

Fig.\ref{fig4} shows how well the above formula fits the eigenvalue 
distribution of reduced density matrices corresponding to the eigenstates of
the coupled tops. Time evolving states also have the same distribution.
This figure also shows that the probability of getting an eigenvalue outside 
the range $[\lambda_{\mbox{\small min}}, \lambda_{\mbox{\small max}}]$ is 
indeed very small. The sum in $S_V$ [see Eq.(\ref{Sv})] can be replaced by 
an integral over the density $f(\lambda)$:
\be
S_V \sim - N \int_{\lambda_{\mbox{\small min}}}^{\lambda_{\mbox{\small max}}}
\,f(\lambda) \lambda \ln \lambda\,d\lambda \equiv \ln(\gamma N) = \ln(N) +
\ln(\gamma).
\ee
\n The integral in $\gamma$ can be evaluated to a generalized hypergeometric 
function and the final result is
\be
\gamma = \frac{Q}{Q+1} \exp\left[\frac{Q}{2(Q+1)^2} ~{{_3}F_{2}} \left\{
1, 1, \frac{3}{2}; 2, 3; \frac{4Q}{(Q+1)^2} \right\} \right].
\ee
\n When the Hilbert space dimension of the subsystems are equal, that is $Q=1$,
the above expression gives $\gamma = \exp(-\frac{1}{2})$, and therefore the 
corresponding $S_V$ is $\ln(N) - \frac{1}{2}$. This is the statistical bound on
entanglement found in our numerical study. In another extreme case, when 
$M \gg N$, that is $Q \gg 1$, then $\gamma \sim 1$ and the corresponding $S_V$
is equal to its maximum possible value $[\ln(N)]$. Therefore, the analytical
formulation based on RMT is able to explain the saturation behavior of quantum
entanglement production very accurately. 

\subsection{Entanglement production in coupled strongly chaotic system}

We have already mentioned that, due to the simpler form of the linear entropy
$S_R$, it is easier to derive an approximate formula for its time evolution.
Here we now present an analytical formalism for the time evolution of $S_R$
in coupled strongly chaotic systems. Let us start the formalism with the 
assumption that the initial state is a product state, given as $|\psi(0)\rangle
= |\phi_1(0)\rangle \otimes |\phi_2(0)\rangle$, where $|\phi_i(0)\rangle$'s 
are the states corresponding to individual subsystems. In general, the time
evolution operator of a coupled system is of the form $U = U_{\epsilon} U_0
= U_{\epsilon} (U_1 \otimes U_2)$, where $U_{\epsilon}$ is the coupling time
evolution operator and $U_i$'s are the time evolution operators of the 
individual subsystems. Furthermore, we have assumed $U_{\epsilon} =
\exp(- i \epsilon H_{12})$ where $H_{12} = h^{(1)} \otimes h^{(2)}$, and 
$h^{(i)}$ are Hermitian local operators. Here we derive our formalism in the
eigenbasis of $h^{(i)}$'s, i.e., $h^{(i)}|e_{\alpha}^{(i)}\rangle = 
e_{\alpha}^{(i)} |e_{\alpha}^{(i)}\rangle$, where $\{e_{\alpha}^{(i)}, 
|e_{\alpha}^{(i)}\rangle\}$ are the eigenvalues and the corresponding 
eigenvectors of $h^{(i)}$.

The one step operation of $U$ on $|\psi(0)\rangle$ will give the time evolving
state at time $n = 1$, i.e., $|\psi(1)\rangle$. Now, at this time, we can 
determine the matrix elements of RDM corresponding to the first subsystem in 
the eigenbasis of $h^{(i)}$'s as 
\be
\hspace{-1.5cm}
[\rho_1(1)]_{\alpha\beta} = \sum_{\gamma} \exp\left[- i \epsilon \left(
e_{\alpha}^{(1)} - e_{\beta}^{(1)}\right) e_{\gamma}^{(2)}\right] \langle 
e_{\alpha}^{(1)}, e_{\gamma}^{(2)} | \psi_0(1)\rangle \langle \psi_0(1) | 
e_{\beta}^{(1)}, e_{\gamma}^{(2)}
\rangle,
\ee
\n where $|\psi_0(1)\rangle$ is the time evolving state of the uncoupled 
system. We now assume that $|\psi_0(1)\rangle$ is a random vector, and 
consequently we can further assume that the components of $|\psi_0(1)\rangle$
are uncorrelated to the exponential term coming due to the coupling. Therefore,
we have
\be
[\rho_1(1)]_{\alpha\beta} \simeq \frac{1}{N}\,\bigl[ \rho_{10}(1) 
\bigr]_{\alpha\beta}\, \sum_{\gamma}\, \exp\left[-i\epsilon 
\left(e_{\alpha}^{(1)}-e_{\beta}^{(1)}\right)\,e_{\gamma}^{(2)}\right],
\ee
\begin{figure}[t]
\epsfxsize=7.5cm
\centerline{\epsfbox{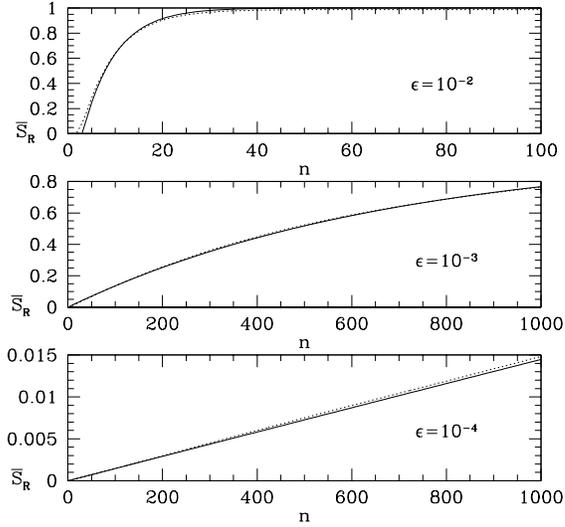}}
\caption{Evolution of the Linear entropy for the coupled strongly chaotic
system is presented. The dotted line is the numerical results of the coupled
kicked tops system. We choose $k=6.0$ for the first top and $k = 6.1$ for
the second top. The solid line is the theoretical estimation, given by Eq.}
\label{fig5}
\end{figure}
\n where $N$ is the Hilbert space dimension of the first subsystem and
$\rho_{10}$ is the density matrix corresponding to the uncoupled top. If we
follow same procedure for one more time step, then at $n=2$ we have
\bmath
\be
\bigl[ \rho_1(2) \bigr]_{\alpha\beta} \simeq \frac{1}{N} |p(\epsilon)|^2
\bigl[ \rho_{10}(2) \bigr]_{\alpha\beta} \sum_{\gamma}\,
\exp\left[-i\epsilon\left(e_{\alpha}^{(1)} - e_{\beta}^{(1)}\right)\right]
\ee
\n where
\be 
p(\epsilon) = \frac{1}{N^2} \sum_{\alpha,\beta}
\exp\left(-i \epsilon\, e_{\alpha}^{(1)}\, e_{\beta}^{(2)}\right).
\ee
\emath
\n If we evolve the initial state for any arbitrary time $n$ and follow the
similar procedure as described above, we get
\be
\hspace{-1cm}
\bigl[ \rho_1(n) \bigr]_{\alpha\beta} = \frac{1}{N}
|p(\epsilon)|^{2(n-1)} \bigl[ \rho_{10}(n) \bigr]_{\alpha\beta}
\sum_{\gamma}\,\exp\left[-i\epsilon\left( e_{\alpha}^{(1)}- e_{\beta}^{(1)}
\right)\,e_{\gamma}^{(2)}\right]. 
\ee
\n It is now straightforward to calculate linear entropy from the above 
expression, and that is given as
\be
\hspace{-1cm}
S_R(n) \simeq 1-\frac{1}{N^4}|p(\epsilon)|^{4(n-1)} \sum_{\alpha,\beta}
\sum_{\gamma,\delta}\exp\left[-i\epsilon\left(e_{\alpha}^{(1)}-e_{\beta}^{(1)}
\right)\left(e_{\gamma}^{(2)}-e_{\delta}^{(2)}\right)\right].
\ee
\n For the coupled kicked tops $H_{12} = J_{z_1} \otimes J_{z_2}/j$. Therefore,
for this particular system, the above formula would become in large $j$-limit
as
\bmath
\bearr
S_R(n) &\simeq& 1 - p(\epsilon)^{4(n-1)} \Biggl[ \frac{2}{N}\left\{ 1 +
\frac{\mbox{Si}\bigl( 2N\epsilon \bigr)}{\epsilon}\right\}-
\left(\frac{1}{N\epsilon} \right)^2 \nonumber\\ 
&&\times\bigl\{ 1 - \cos \bigl( 2N\epsilon\bigr)
\bigr. \Biggr. 
+ \mbox{Ci}\bigl( 2 N \epsilon \bigr) \,-\, \ln\bigl( 2 N \epsilon \bigr)
- \gamma \Biggl. \bigl. \bigr\}\Biggr]
\eearr
\n where
\be
p(\epsilon) \simeq \frac{2}{N} \left[ 1 + \frac{1}{\epsilon} \mbox{Si}
\left(\frac{N \epsilon}{2}\right)\right].
\ee
\emath 
\n The functions Si and Ci are the standard {\it sine-integral} and 
{\it cosine-integral} function, respectively, while $\gamma = 0.577216...$
is the Euler constant. In the above formulation we have not assumed, unlike
the perturbation theory \cite{tanaka}, any particular order of magnitude of
the coupling strength $\epsilon$.  

In Fig.\ref{fig5}, we have presented our numerical result of the linear entropy
$(S_R)$ production in the coupled tops where the individual tops are strongly
chaotic. In the above formalism, we have not assumed any special symmetry
property. Here we break permutation symmetry by taking two nonidentical tops
with $k = 6.0$ for the first top and $k = 6.1$ for the second. Fig.\ref{fig5}
demonstrates that how well our theoretical estimation, denoted by the 
solid curve, is valid for both weak and strong coupling strengths.    

\section{\label{sec5}Summary}

In this paper, our major goal was to study entanglement production in coupled
chaotic system as a possible quantum indicator of classical chaos. We have 
used coupled kicked top as a model for our study. Single kicked top is a well 
studied model of both classical and quantum chaotic system. Therefore, it is 
easier for us to identify the effect of underlying classical dynamics on the 
entanglement production. We have studied entanglement production for different 
underlying classical dynamics of the individual top and also for different 
coupling strengths. Here we not only have considered entanglement production 
for the pure initial state, but we have also initiated the study of 
entanglement production in coupled chaotic systems for the mixed initial state.
We have used Log-negativity, a recently proposed measure, to characterize 
mixed state entanglement production. In general, for both kind of initial 
states, entanglement production is higher for stronger chaotic cases. However, 
we have observed a saturation of entanglement production when individual tops 
are strongly chaotic and they are coupled strongly to each other. Coupling 
strength between two tops is also a crucial parameter for the entanglement 
production. For instance, when the coupling strength between two tops is very 
weak, we find higher entanglement production for sufficiently long time 
corresponding to non-chaotic cases. This is also a common property of both 
kind of initial states. We analytically estimated the above mentioned 
statistical bound on the pure state entanglement (saturation of entanglement 
production) using RMT. We have also derived an approximate formula, based again
on the ideas of RMT, for the pure state entanglement production in coupled 
strongly chaotic system. This formula is applicable, unlike perturbation 
theory, to large coupling strengths and is also valid for sufficiently long 
time. We still do not have deeper analytical understanding of mixed state
entanglement production in chaotic systems, an open problem that is related
to the establishment of calculable measures of mixed state entanglement.

\end{document}